# Early differentiation of magmatic iron meteorite parent bodies from Mn–Cr chronometry

A. Anand[1*], J. Pape[1,2], M. Wille[1], K. Mezger[1], B. Hofmann[1,3]

[1] Institut für Geologie, Universität Bern, Baltzerstrasse 1+3, 3012 Bern, Switzerland

[2] Institut für Planetologie, Universität Münster, Wilhelm-Klemm-Str. 10, 48149 Münster, Germany

[3] Naturhistorisches Museum Bern, Bernastrasse 15, CH-3005, Bern, Switzerland.

*Corresponding author (email: aryavart.anand@geo.unibe.ch)

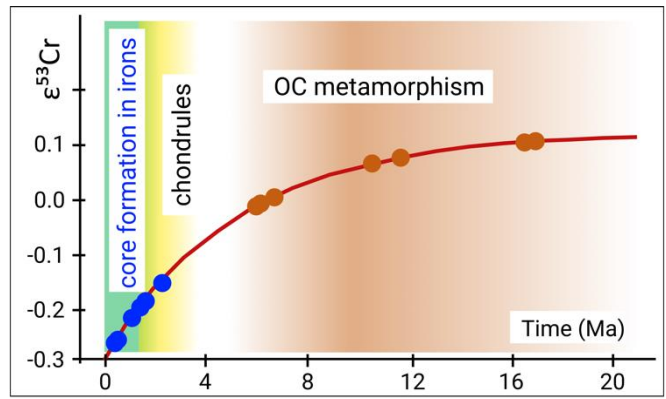

**Abstract**

Magmatic iron meteorite groups such as IIAB, IIIAB and IVA, represent the largest sampling of extraterrestrial core material from the earliest accreted distinct planetary bodies in the solar system. Chromium isotope compositions of chromite/daubréelite from seven samples, translated into $^{53}Cr/^{52}Cr$ model ages, provide robust time information on planetary core formation. These ages range within ~1.5 Ma after formation of calcium-aluminium-rich inclusions (CAIs) and define the time of metal core formation in the respective parent bodies, assuming metal–silicate separation was an instantaneous event that induced strong chemical fractionation of Mn from the more siderophile Cr. The early core formation ages support accretion and differentiation of the magmatic iron meteorite parent bodies to have occurred prior to the chondrule formation interval. The calibration of Mn–Cr ages with established Hf–W ages of samples from the same magmatic iron meteorite groups constrains the initial $\varepsilon^{53}Cr$ of the solar system to −0.30 ± 0.05, and thus lower than previously estimated.

**Introduction**

Members of the different magmatic iron meteorite groups are thought to sample the cores of distinct parent bodies that experienced large scale chemical fractionation, most notably metal–silicate separation. The absolute time of core formation provides a key time marker for the evolution of early formed planetesimals including accretion and cooling of the respective parent body. The most commonly used chronological system to date iron meteorites is the $^{182}Hf$–$^{182}W$ system (Kruijer *et al*., 2017 and references therein), constraining core formation in iron meteorite parent bodies over an interval of ~1 Myr and their accretion to ~0.1–0.3 Ma after the formation of Ca-Al-rich inclusions (4567.18 ± 0.50 Ma; Amelin *et al*., 2010). These early accretion ages predate or are contemporaneous with the chondrule formation interval (*e.g.*, Connelly *et al.*, 2012; Pape *et al.*, 2019). However, correct interpretation of Hf–W data depends on the accurate knowledge of initial $\varepsilon^{182}$W of the solar system and Hf/W ratios of the parent bodies which are well established but still needs to consider possible variations in Hf isotopes due to galactic cosmic radiation (GCR) (Kruijer *et al.*, 2017).

Another powerful tool to constrain the time and duration of early solar system processes, including accretion, differentiation, metamorphism and subsequent cooling could be the short lived $^{53}$Mn–$^{53}$Cr chronometer ($t_{1/2}$ = 3.7 ± 0.4 Ma; Honda and Imamura, 1971) (*e.g.*, Shukolyukov and Lugmair, 2006; Trinquier *et al.*, 2008; Göpel *et al.*, 2015; Zhu *et al.*, 2021). Chromite ($FeCr_2O_4$) and daubréelite ($FeCr_2S_4$) are the two main carrier phases of Cr in magmatic iron meteorites. Both minerals have low Mn/Cr ratios (≤0.01; Duan and Regelous, 2014) and thus preserve the Cr isotope composition of their growth environment at the time of isotopic closure, while the in-growth of radiogenic $^{53}$Cr from *in situ* decay of $^{53}$Mn is negligible (Anand *et al.*, 2021). This makes them suitable for obtaining model ages by comparing their Cr isotopic composition with the Cr isotope evolution of the host reservoir. A particular advantage is that low Fe/Cr ratios in chromite and daubréelite (typically ~0.5), result in negligible contribution of $^{53}$Cr produced by GCR from Fe; hence no correction for spallogenic Cr is required (Trinquier *et al.*, 2008; Liu *et al.*, 2019) (Supplementary Information).

This study presents model ages for chromite and daubréelite from the largest magmatic iron meteorite group collections (IIAB, IIIAB and IVA) that constrain the earliest stages of planetesimal formation and differentiation. These Cr model ages define the timing of metal segregation during core formation. Chromium-rich phases formed in the metal inherit the Cr isotope composition of their low Mn/Cr host and thus constrain the time of last silicate–metal equilibration.

**Methods**

A chromite or daubréelite fraction from seven iron meteorites was analysed. After mineral digestion and chemical purification, Cr isotopes were measured on a Triton™ Plus TIMS at the University of Bern. Each sample was measured on multiple filaments to achieve high precision for $^{53}$Cr/$^{52}$Cr ratio. Isotope compositions are reported as parts *per* 10,000 deviations (ε notation) from the mean value of a terrestrial Cr standard measured along with the samples in each session. External precision (2 s.d.) for the terrestrial standard in a typical measurement session was ±0.1 for $\varepsilon^{53}$Cr and ±0.2 for $\varepsilon^{54}$Cr (Supplementary Information).

***Model for ε⁵³Cr evolution in chondritic reservoir***. Model $^{53}$Cr/$^{52}$Cr ages for early formed solar system bodies and their components can be determined on materials with high Cr/Mn and considering the following (i) homogeneous distribution of $^{53}$Mn in the solar system (*e.g.*, Trinquier *et al*., 2008; Zhu *et al*., 2019), (ii) known abundances of $^{53}$Mn and $^{53}$Cr at the beginning of the solar system (*i.e.* solar system initial ε$^{53}$Cr) or any point in time thereafter, (iii) an estimate for the Mn/Cr in the relevant reservoir, and (iv) known decay constant of $^{53}$Mn. Based on these assumptions the evolution of the $^{53}$Cr/$^{52}$Cr isotope composition of the chondritic reservoir through time can be expressed as:

$$(^{53}Cr/^{52}Cr)_p = (^{53}Cr/^{52}Cr)_i + (k)(^{53}Mn/^{55}Mn)_i \times (1 - e^{-\lambda t}) \qquad \text{Eq. 1}$$

where the subscripts 'p' and 'i' refer to the present day and initial solar system values, respectively, and $\lambda$ denotes the $^{53}$Mn decay constant. The $^{55}$Mn/$^{52}$Cr of the reservoir is denoted by *k*; and *t* represents the time elapsed since the start of the solar system, which is equated with the time of formation of CAIs. Equation 1 describes the evolution of $^{53}$Cr/$^{52}$Cr with time for the chondritic reservoir and can be used to derive model ages for a meteorite sample by measuring the Cr isotopic composition of its chromite/daubréelite fraction.

**Results**

The ε$^{53}$Cr and ε$^{54}$Cr of chromite/daubréelite fractions determined for all samples are listed in Table 1. No correlation is observed in ε$^{53}$Cr *vs*. ε$^{54}$Cr and ε$^{53}$Cr *vs*. Fe/Cr that corroborates an insignificant spallogenic contribution (Supplementary Information). Model ages are calculated relative to the CAI formation age of 4567.18 ± 0.50 Ma (Amelin *et al.,* 2010) assuming an OC chondritic $^{55}$Mn/$^{52}$Cr = 0.74 (Zhu *et al.,* 2021), a solar system initial ε$^{53}$Cr = −0.23 and a canonical $^{53}$Mn/$^{55}$Mn = 6.28 × 10$^{-6}$ (Trinquier *et al.,* 2008) (Fig. 1).

**Table 1** Mn/Cr, Fe/Cr and Cr isotopic compositions of chromite and daubréelite fractions from iron meteorites.

| Sample (Coll. Number) | Group/Fraction | Mn/Cr | Fe/Cr | $\varepsilon^{53}$Cr | 2 s.e. | $\varepsilon^{54}$Cr | 2 s.e. | Model Age[a] | Model Age[b] | $n$ |
|---|---|---|---|---|---|---|---|---|---|---|
| Agoudal (43830) | IIAB/Chromite | 0.0053(4) | 0.37(1) | −0.210 | 0.023 | −0.784 | 0.060 | $0.27^{+0.33}_{-0.31}$ | $1.33^{+0.41}_{-0.38}$ | 12 |
| Sikhote Alin (43380) | IIAB/Chromite | 0.0052(8) | 0.41(3) | −0.228 | 0.025 | −0.923 | 0.051 | $0.03^{+0.34}_{-0.32}$ | $1.04^{+0.41}_{-0.38}$ | 12 |
| NWA 11420 (43837) | IIAB/Daubréelite | 0.0040(1) | 0.59(1) | −0.203 | 0.045 | −0.768 | 0.055 | $0.37^{+0.67}_{-0.59}$ | $1.45^{+0.83}_{-0.72}$ | 7 |
| Saint Aubin[c] | IIIAB/Chromite | 0.0096(2) | 0.57(1) | −0.268 | 0.029 | −0.779 | 0.061 | $-0.47^{+0.35}_{-0.33}$ | $0.44^{+0.42}_{-0.39}$ | 10 |
| Cape York (33137) | IIIAB/Chromite | 0.0062(3) | 0.46(1) | −0.196 | 0.043 | −0.780 | 0.062 | $0.47^{+0.64}_{-0.57}$ | $1.57^{+0.80}_{-0.70}$ | 13 |
| Yanhuitlan (50084) | IVA/Daubréelite | 0.0490(7) | 0.54(1) | −0.272 | 0.027 | −0.468 | 0.061 | $-0.52^{+0.33}_{-0.31}$ | $0.38^{+0.39}_{-0.36}$ | 10 |
| Duchesne (50033) | IVA/Chromite | 0.0048(4) | 1.1(1) | −0.160 | 0.037 | −0.487 | 0.156 | $1.00^{+0.61}_{-0.55}$ | $2.23^{+0.78}_{-0.68}$ | 7 |
| IAG OKUM | whole rock std. | 0.60(6) | 36.6(9) | +0.020 | 0.065 | +0.083 | 0.110 | | | 7 |

Collection numbers refer to meteorite collections of NHM Bern.

The uncertainties associated with Mn/Cr, Fe/Cr, and Cr isotopic compositions are reported as 2 s.e. of the replicate measurements. See Table S-1 for Cr isotopic composition of the individual runs. $n$ = number of replicate measurements.

[a] Equation 1, $\varepsilon^{53}Cr_i = -0.23$.

[b] Equation 1, $\varepsilon^{53}Cr_i = -0.30$.

[c] NHM Vienna collection ID for Saint Aubin: NHMV_#13635_[A].

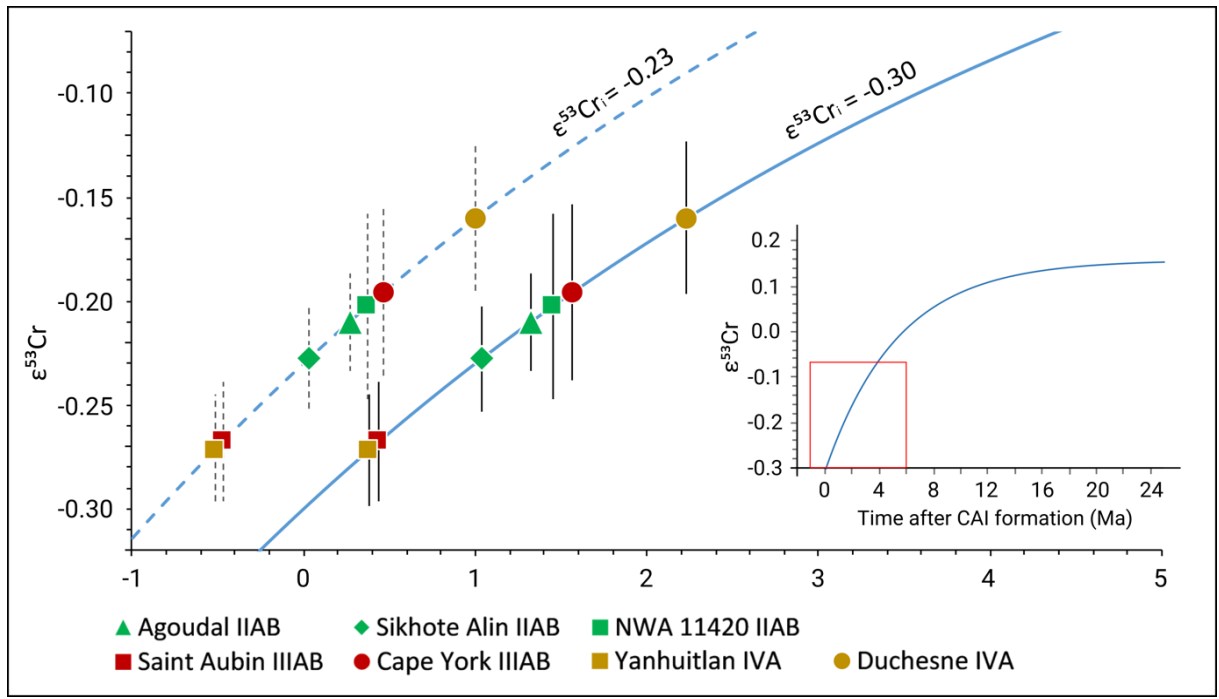

**Figure 1**     ε$^{53}$Cr values of chromite/daubréelite plotted on a Cr isotope evolution curve determined for a chondritic reservoir through time using Equation 1. Error bars represent 2 s.e. uncertainties.

**Discussion**

The inference of Mn–Cr model ages to date core formation events is based on the assumption that metal–silicate separation was instantaneous. It occurred when the chondritic parent bodies of magmatic iron meteorites were heated by accretion energy and the decay of short lived $^{26}$Al, and reached the liquidus temperature of iron–sulfur alloy (1325 °C to 1615 °C, depending on the S content in the metal melt; Kaminski *et al*., 2020 and references therein). The metal–silicate separation induced a strong chemical fractionation of Mn from the more siderophile Cr (Mann *et al*., 2009). The measured low Mn/Cr (≤0.01; Duan and Regelous, 2014) in iron meteorites corroborates the efficiency of this fractionation. Because of the low Mn/Cr of the metallic core, its Cr isotopic composition remained unchanged and reflects the composition at the time of metal–silicate differentiation. Therefore, the Cr isotopic composition of chromite/daubréelite that formed in the metallic core, no matter at what time after the metal segregation, reflects the time of Mn/Cr fractionation from a reservoir with CI chondritic $^{55}$Mn/$^{52}$Cr and not the time of mineral formation nor its closure below a certain closure temperature.

The Mn–Cr model ages determined using Equation 1. assume a Mn/Cr for the source reservoir that is represented by the average Mn/Cr of OCs. Ordinary chondrites have a CI-like Mn/Cr (*e.g.*, Wasson and Allemeyn, 1988; Zhu *et al.*, 2021) and alongside the investigated magmatic iron meteorite groups, belong to the 'non-carbonaceous' reservoir (Kleine *et al.*, 2020). The effect of different Mn/Cr of the iron meteorite parent bodies and the assumption of different initial $\varepsilon^{53}$Cr is shown in Figure S-2. Model ages for IIAB, IIIAB and IVA groups are unaffected by the growth trajectory chosen, given current analytical resolution. Assuming a Mn/Cr similar to carbonaceous chondrites would change the model ages by a maximum of 1 Ma for the youngest sample. However, since all samples belong to the non-carbonaceous group, the average composition of OCs is most appropriate.

Since the Cr isotopic composition of the samples is unaffected by contributions from spallogenic Cr (Supplementary Information), the only major source of uncertainty in Mn–Cr model ages comes from the choice of initial $\varepsilon^{53}$Cr and $^{53}$Mn/$^{55}$Mn values. Figure S-3 shows Mn–Cr model ages for the studied samples determined using initial Mn–Cr isotopic compositions from multiple studies reporting resolvable variations in the solar system initial $\varepsilon^{53}$Cr and $^{53}$Mn/$^{55}$Mn values. Clearly, more high precision Cr isotope data for samples dated with different chronometers are needed to further constrain the initial $\varepsilon^{53}$Cr and $^{53}$Mn/$^{55}$Mn. Model ages determined using initial $\varepsilon^{53}$Cr and $^{53}$Mn/$^{55}$Mn values from Göpel *et al.* (2015) and Shukolyukov and Lugmair (2006) predate the CAI formation age, contradicting the standard solar system model in which CAIs are the earliest formed solid objects. The Mn–Cr model ages determined using initial $\varepsilon^{53}$Cr and $^{53}$Mn/$^{55}$Mn from Trinquier *et al.* (2008) mostly postdate CAI formation and thus appear generally more reliable.

The Mn–Cr model ages can also be compared with other chronometers that have been used to date meteorites and their components. $^{182}$Hf–$^{182}$W, $^{207}$Pb–$^{206}$Pb and $^{187}$Re–$^{187}$Os are some of the common chronological systems providing constraints on different stages in the evolution of iron meteorite parent bodies (Goldstein *et al.*, 2009 and references therein). However, when applied to iron meteorites all other chronological systems date cooling below their respective isotopic closure with the exception of the $^{182}$Hf–$^{182}$W system, which has strong similarities to the $^{53}$Mn–$^{53}$Cr system. It is also suitable for examining the timescales and mechanisms of metal segregation for iron meteorite parent bodies since Hf and W have different geochemical behaviours resulting in strong Hf/W fractionation during metal/silicate separation (*i.e.* core formation). However, in addition to the uncertainty on the initial $\varepsilon^{182}$W of the solar system, $^{182}$W/$^{184}$W data are also affected by secondary neutron capture effects on W isotopes induced during cosmic ray exposure (unlike Mn–Cr model ages reported here).

Recently, Pt isotope data have been used to quantify the effects of neutron capture on W isotope compositions, making it possible to produce more reliable core formation ages (*e.g.*, Kruijer *et al.*, 2017). The Mn–Cr core formation age corresponding to the weighted mean $\varepsilon^{53}$Cr of combined IIAB, IIIAB and IVA groups determined using solar system initial $\varepsilon^{53}$Cr = −0.23 (Trinquier *et al.*, 2008) is ~1 Myr older than the Hf–W core formation age corresponding to Pt corrected weighted mean $\varepsilon^{182}$W of the same iron groups (Fig. 2, Table S-2) (Kruijer *et al.*, 2017). However, the Hf–W and Mn–Cr systems show consistent crystallisation ages in angrites (internal isochrons established by minerals) that also belong to the 'non-carbonaceous' reservoir and originated from differentiated parent bodies (Zhu *et al.*, 2019). The different chronometers are expected to agree because of the rapid cooling of angrites indicated by their basaltic texture. A better fit between Hf–W and Cr model ages can be obtained when the uncertainties on the model parameters for Mn–Cr model age determination are considered. Uncertainties on the $^{53}$Mn decay constant (Honda and Imamura, 1971) and solar system $^{53}$Mn/$^{55}$Mn (Trinquier *et al.*, 2008) result in only a minor shift in the model ages of generally <0.02 Myr which is insignificant. However, using a solar system initial $\varepsilon^{53}$Cr = −0.30, which is within its reported uncertainty ($\varepsilon^{53}$Cr = −0.23 ± 0.09; Trinquier *et al.*, 2008), results in a perfect fit with the mean $^{182}$Hf–$^{182}$W model ages for magmatic iron meteorite groups (Fig. 2). Consequently, $\varepsilon^{53}$Cr = −0.30 is proposed as a better estimate for the solar system initial $\varepsilon^{53}$Cr. To maintain a match between Hf–W and Mn–Cr model ages the uncertainty on the initial $\varepsilon^{53}$Cr of the solar system is less than ±0.05.

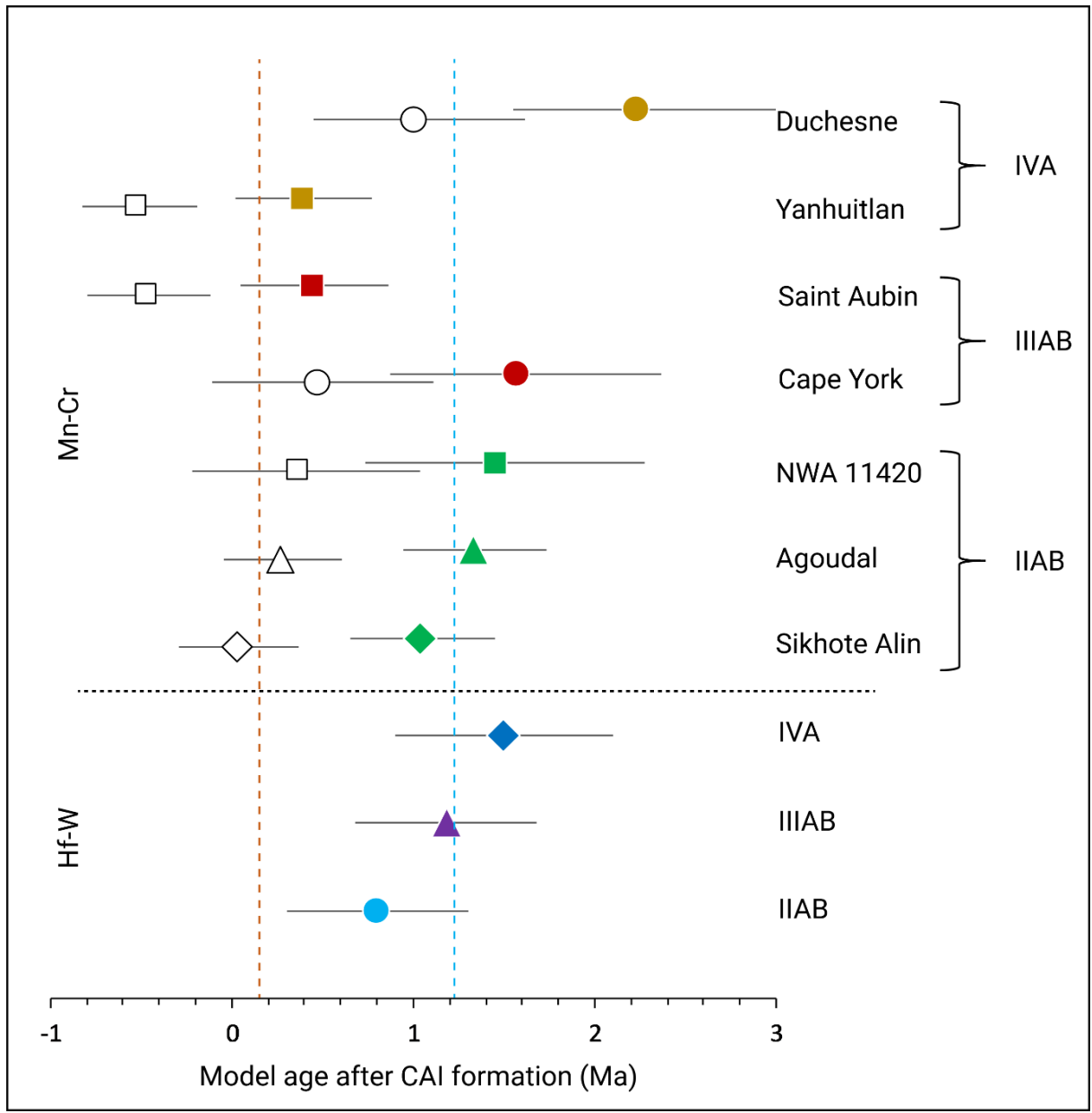

**Figure 2** Comparison between Mn–Cr (present study) and Hf–W (Kruijer *et al.*, 2017) core formation ages. The Mn–Cr ages are determined using Equation 1 and $\varepsilon^{53}Cr_i = -0.23$ (open symbols) from Trinquier *et al.* (2008) and $\varepsilon^{53}Cr_i = -0.30$ (filled symbols) proposed in the present study.

Figure 3 presents a timeline depicting chromite/daubréelite model ages for IIAB, IIIAB and IVA iron meteorites and parent body metamorphism ages for type 3 and 6 ordinary chondrites as determined in Anand *et al.* (2021) using updated parameters for model age calculation. Combined with the existing thermal models (*e.g.*, Qin *et al.*, 2008), Hf–W core formation ages and calibrated Mn–Cr model ages constrain the accretion of the magmatic iron meteorite parent bodies to within less than 1 Myr and no later than 1.5 Myr after CAI formation. This is in perfect agreement with numerical simulations that require early and efficient accretion of larger bodies within the protoplanetary disk (*e.g.*, Johansen *et al*., 2007; Cuzzi *et al*., 2008). The small spread in the model ages of samples from the same meteorite group might reflect some core–mantle exchange during the solidification of the metal core. The range is similar to the range of individual Hf–W model ages within an iron-meteorite group (*e.g.*, Kruijer *et al*., 2017).

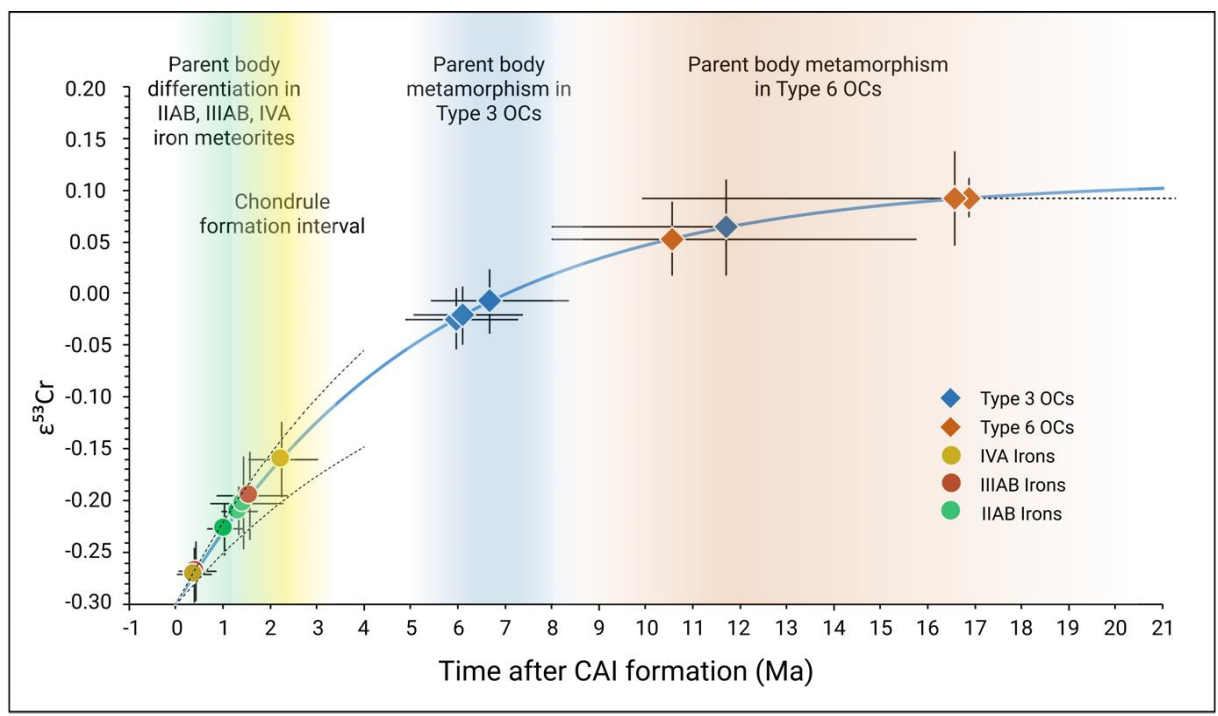

**Figure 3** Timeline of early solar system formation showing parent body differentiation in magmatic iron meteorites, chondrule formation, and parent body metamorphism on ordinary chondrites (see text for references). Error envelope over iron meteorites shown by dashed lines represent the maximum variation in the evolutionary paths due to different Mn/Cr ratios of the parent bodies (see Fig. S-2).

One of the most important implications of Mn–Cr and Hf–W (*e.g.*, Kruijer *et al*., 2017; Spitzer *et al*., 2021) core formation ages is that they bring the accretion and differentiation of the magmatic iron meteorite parent bodies in context with the chondrule formation interval recorded in chondrite samples (*e.g.*, Connelly *et al*., 2012; Pape *et al*., 2019, 2021). $^{207}$Pb–$^{206}$Pb chondrule formation ages (Connelly *et al*., 2012) suggest that the production of chondrules began as early as the CAI condensation; hence, contemporaneous with the accretion of the parent bodies of magmatic iron meteorites as suggested by Hf–W core formation ages and collaborated by Mn–Cr model ages in the present study. $^{26}$Al–$^{26}$Mg ages for the formation of melt in individual chondrules, as summarised in Pape *et al.* (2019), suggest that chondrule formation in ordinary and most carbonaceous chondrites lasted from *ca*. 1.8–3.0 Ma with a major phase around 2.0–2.3 Ma after CAI formation. This puts the chondrule formation interval after the accretion of the magmatic iron meteorite parent bodies. The latter implies that chondrule formation may not necessarily be an intermediate step on the way from dust to planets, but rather early planet formation may have been the cause for the chondrule formation at least in extant chondrite samples. Thus, the early planetesimal formation (*i.e.* accretion of the iron meteorites parent bodies) was a local process and happened while other regions were still mostly in the stage of accreting dust particles and chondrule formation.

## Acknowledgements

We appreciate the support through 'Swiss Government Excellence Scholarship (2018.0371)' and NCCR PlanetS supported by the Swiss National Science Foundation grant no. 51NF40-141881. We thank Dr. Ludovic Ferriere from NHM Vienna for providing chromite from Saint Aubin meteorite. Dr. Harry Becker and Smithsonian Institution are thanked for providing Allende powder sample. Patrick Neuhaus and Lorenz Gfeller from the Institute of Geography, University of Bern are thanked for assistance with the ICP-MS analysis of the samples. We thank Dr. Maud Boyet for editorial handling and Dr. Ke Zhu and an anonymous reviewer for their constructive comments that helped to improve the manuscript.